\documentclass[aps,pra,twocolumn,notitlepage,superscriptaddress,showpacs,longbibliography,10pt]{revtex4-1}

\usepackage{amsmath}    
\usepackage{amsfonts}
\usepackage{bm}
\usepackage{amssymb}
\usepackage{appendix}
\usepackage{graphicx}   
\usepackage{color}      
\usepackage{subfigure}  
\usepackage{comment}
\usepackage{siunitx}
\usepackage{color,soul,url}

\usepackage{soul}
\usepackage[normalem]{ulem}

\newcommand{\re}{\textrm{Re}}
\newcommand{\im}{\textrm{Im}}

\begin{document}

\title{Bound states in the continuum through environmental design}
\author{Alexander Cerjan}
\email[]{awc19@psu.edu}
\affiliation{Department of Physics, The Pennsylvania State University, University Park, Pennsylvania 16802, USA}
\author{Chia Wei Hsu}
\affiliation{Ming Hsieh Department of Electrical Engineering, University of Southern California, Los Angeles, California 90089, USA}
\author{Mikael C. Rechtsman}
\affiliation{Department of Physics, The Pennsylvania State University, University Park, Pennsylvania 16802, USA}

\date{\today}

\begin{abstract}
  We propose a new paradigm for realizing bound states in the continuum (BICs) by  engineering the environment of a system to control the number of
  available radiation channels. Using this method, we demonstrate that
  a photonic crystal slab embedded in a photonic crystal environment can exhibit both isolated points and lines of BICs in
  different regions of its Brillouin zone. Finally, we demonstrate that the intersection between a line of BICs and line of leaky resonance can
  yield exceptional points connected by a bulk Fermi arc. The ability to design the environment of a system
  opens up a broad range of experimental possibilities for realizing BICs in three-dimensional geometries, such
  as in 3D-printed structures and the planar grain boundaries of self-assembled systems.
\end{abstract}

\maketitle

Bound states in the continuum (BICs), which are radiation-less states in an open system whose frequency
resides within the band of radiative channels, have recently attracted a great deal of interest for
their applications in producing vector beams from surface emitting lasers \cite{meier_laser_1998,imada_coherent_1999,noda_polarization_2001,miyai_photonics:_2006,matsubara_gan_2008,iwahashi_higher-order_2011,kitamura_focusing_2012,hirose_watt-class_2014}
and enhancing the resolution
of certain classes of sensors \cite{yanik_seeing_2011,zhen_enabling_2013,romano_label-free_2018}. Originally proposed in 1929 in a quantum mechanical context \cite{original_1929}, BICs have now
been found in a
broad range of physical systems, such as photonic crystal slabs \cite{paddon_two-dimensional_2000,pacradouni_photonic_2000,ochiai_dispersion_2001,fan_analysis_2002,hsu_bloch_2013,hsu_observation_2013,yang_analytical_2014,zhen_topological_2014,zhou_perfect_2016,gao_formation_2016,kodigala_lasing_2017,zhang_extraordinary_2018,minkov_zero-index_2018}, waveguide arrays \cite{plotnik_experimental_2011,weimann_compact_2013,corrielli_observation_2013},
strongly coupled plasmonic-photonic systems \cite{azzam_formation_2018}, metasurfaces \cite{koshelev_meta-optics_2018}
acoustics \cite{parker_1966,parker_1967,cumpsty_1971,koch_1983,parker_excitation_1989,evans_existence_1994},
and water waves \cite{ursell_trapping_1951,jones_eigenvalues_1953,callan_trapped_1991,retzler_trapped_2001,cobelli_experimental_2009,cobelli_experimental_2011}.
Additionally, lines of BICs were recently found in composite birefringent structures \cite{gomis-bresco_anisotropy-induced_2017,mukherjee_topological_2018}.
In principle, BICs can be
classified into three main categories \cite{hsu_bound_2016}: those which are engineered using an inverse construction
method, those which are protected by symmetry or separability, and those which can be found `accidentally'
through tuning a system's parameters. In practice however, systems supporting BICs from the first category are difficult to experimentally
realize due to the high degree of fine-tuning required. Thus, much of the current excitement surrounding BICs has focused on systems which feature symmetry-protected and accidental BICs; moreover, these BICs have been shown to possess topological protection that guarantees their existence under perturbations to the system \cite{zhen_topological_2014,bulgakov_topological_2017,xiao_topological_2017,zhang_observation_2018,doeleman_experimental_2018,takeichi_topological_2018}.


Traditionally, the appearance of
accidental BICs is understood in terms of modal interference \cite{friedrich_interfering_1985,hsu_bound_2016,gao_formation_2016}, with two or more resonances of the device destructively
interfering in the system's radiation channels and resulting in a bound mode spatially localized to the device. This interpretation emphasizes
how tuning the device's parameters changes the spatial profile of its resonances to realize this modal interference,
while considering the available radiation channels in the surrounding environment as fixed. This is because most previously studied systems
with accidental BICs consider devices embedded in free space, where the outgoing propagating channels
cannot be readily altered. 
However, from this argument it is clear that the environment
is also important in determining the presence or absence of BICs: the environment's properties
dictate the number and modal profiles of the available radiation channels, and thus strongly constrain
when it is possible to achieve the necessary modal interference. Yet thus far, the role of the environment
in creating BICs has remained relatively unexplored.

In this Letter, we show that the properties of the environment play an important role in whether, and where, BICs
can exist, regardless of the specific geometry of the device embedded in this environment. This argument is presented using
temporal coupled-mode theory (TCMT) \cite{haus_waves_1983,suh_temporal_2004,alpeggiani_quasinormal-mode_2017} and as such is completely general and applicable to all systems which exhibit BICs.
As an example of this theory, we will then show that by embedding a photonic crystal slab into a photonic crystal
environment, both isolated BICs and lines of BICs can be found in the resonance bands of the photonic crystal slab depending
on the number of available radiation channels. Moreover, perturbations to the environment can shift the locations
of the system's BICs even when the photonic crystal slab layer remains unchanged, demonstrating that the environment of a system
is an equal partner to the embedded device in determining the existence and types of BICs found in the system.
Finally, we show that when two resonance bands of the photonic crystal slab undergo a symmetry protected band crossing,
it is possible for a line of BICs to pass from one band to the other through a bulk Fermi arc. Understanding the relationship
between the device and surrounding environment in forming BICs is a necessary first step towards realizing BICs in three-dimensional
geometries, such as grain boundaries in photonic crystals \cite{deubel_direct_2004} or self-assembled structures \cite{blanco_large-scale_2000}.

To illuminate the role of the environment in determining the presence and properties of BICs in a system, we first consider
a photonic crystal slab embedded in an environment, such that the entire system is periodic in $x$ and $y$. The dynamics of
an isolated resonance at any choice of in-plane wavevector, $\mathbf{k}_\parallel = (k_x, k_y)$, of the photonic crystal slab can be
described using temporal coupled-mode theory (TCMT) as
\begin{equation}
  \partial_t a = -(i\omega_0 + \gamma)a + K^T s_{\textrm{in}},
\end{equation}
where the resonance, with amplitude $a$, frequency $\omega_0$, and decay rate $\gamma$, couples to both the incoming
channels of the environment, $s_{\textrm{in}}$, via the coupling constants $K$, and the outgoing channels of the environment, $s_{\textrm{out}}$,
via
\begin{equation}
  s_{\textrm{out}} = C s_{\textrm{in}} + D a.
\end{equation}
Here, $C$ represents the direct transmission and reflection through the photonic
crystal slab, and is an $N \times N$ complex matrix, where $N$ is the total number of radiative channels both above and below
the photonic crystal slab at frequency $\omega_0$ and $\mathbf{k}_\parallel$,
while $D$ is an $N \times 1$ complex column vector that denotes the outcoupling of the resonance to each available radiation channel.
Finally, if the system possesses $180^\circ$ rotational symmetry about
the $z$-axis ($C_2$), such that $-\mathbf{k}_\parallel$ is equivalent to $\mathbf{k}_\parallel$,
by time-reversal symmetry $C$ and $D$ can be shown to be related as \cite{suh_temporal_2004}
\begin{equation}
  C D^* = - D. \label{eq:CDD}
\end{equation}

In the language of TCMT, a BIC occurs when $D = 0$, i.e.\ all of the complex outcoupling coefficients of the resonance
to all of the available radiation channels are simultaneously zero. For this to occur accidentally with finite probability, there must
be at least as many degrees of freedom of the system as there are unknown parameters of $D$. Thus, naively one would
expect to require $2N$ degrees of freedom to find BICs, as there initially appear to be $2N$ unknown parameters in $D$.
However, while the properties of the resonances, and thus $D$, of the photonic crystal slab are
dependent upon the specific patterning of the slab, the direct scattering processes contained in $C$ are agnostic
to this patterning, and can instead be considered using a homogeneous dielectric slab \cite{fan_analysis_2002,fan_temporal_2003}.
As such, as $C$ is 
essentially constant for perturbations to the photonic crystal slab, 
the requirement of Eq.\ (\ref{eq:CDD}) represents a set of additional constraints on $D$, halving its
number of unknown parameters. Thus, in general if an entire system is $C_2$ symmetric, one only needs $N$ degrees of freedom
to find accidental BICs.

A second symmetry commonly present in photonic crystal slab systems is mirror symmetry about the $z=0$ plane ($\sigma_z$). Although
this symmetry is not required to find BICs, its presence further reduces the number of free parameters among the components of $D$ as the outcoupling coefficients
$d_m$,$d_n$, forming pairs $d_n = \sigma d_m$ with $\sigma = \pm 1$ depending on whether the resonance is even or odd about $z = 0$ \cite{hsu_polarization_2017}.
Here, $m,n$ correspond to the two radiation channels that are mirror-symmetric partners in the regions of the environment which are
above and below the photonic crystal slab that the resonance of interest is coupled to. Thus, mirror symmetry
both halves the number of unknown parameters in $D$ and also halves the number of independent constraints represented by Eq.\ (\ref{eq:CDD}).

\begin{figure}[t!]
    \centering
    \includegraphics[width=0.45\textwidth]{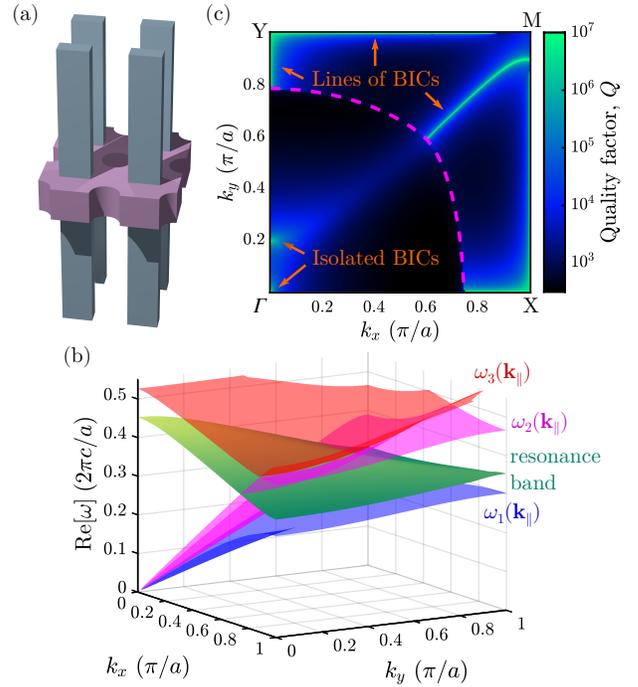}
    \caption{
      (a) Schematic of a photonic crystal slab embedded in a $C_{2v}$ symmetric photonic crystal environment of infinitely tall dielectric rectangles
      (the rectangles in the environment do not penetrate into the slab). The rectangles have length and width
      $l_{\textrm{PhC}} = w_{\textrm{PhC}}/2 = 0.23 a$ and dielectric permittivity $\varepsilon_{\textrm{PhC}} = 12$, while the slab has height $h_{\textrm{slab}} = 0.5 a$,
      holes with radius $r_{\textrm{slab}} = 0.22a$, and dielectric permittivity $\varepsilon_{\textrm{slab}} = 7$,
      where $a$ is the lattice constant of the system. (b) Photonic band structure.
      The band of resonances of the photonic crystal slab are shown in green, while the first three low-frequency cutoffs
      of the radiation channels in the photonic crystal environment, $\omega_n(\mathbf{k}_\parallel)$, are shown in transparent blue, purple, and red, respectively.
      (c) Quality factor of the photonic crystal slab resonances as a function of $\mathbf{k}_\parallel$. Lines of BICs are seen for portions
        of $\Gamma$--$X$, $\Gamma$--$Y$, $X$--$M$, $Y$--$M$, and near $\Gamma$--$M$, where only one environmental channel is present.
      Two isolated BICs are seen at $\Gamma$ and near $k_x = 0$, $k_y = 0.2\pi/a$ where there are two radiative channels. The boundary between the two- and one-radiative-channel regions is marked
      with a dashed purple line.
      Simulations performed using \textsc{Meep} \cite{oskooi_meep:_2010}.
      \label{fig:1}}
\end{figure}

To provide an explicit example of how these constraints can be used to find BICs, consider the photonic crystal slab embedded in a photonic
crystal environment shown in Fig.\ \ref{fig:1}a. The presence of the photonic crystal environment breaks the degeneracy between the two polarizations
of light in homogeneous media, splitting the light line, $\omega = c |\mathbf{k}_\parallel|$, into separate frequency cutoff bands, $\omega_n(\mathbf{k}_\parallel)$,
below which the $n$th radiative channel does not exist \cite{johnson_guided_1999}.
As such, for the TE-like resonance band of the photonic crystal slab shown in Fig.\ \ref{fig:1}b, the 
central region of this band in the Brillouin zone
(below the dashed purple line in Fig.\ \ref{fig:1}c)
can couple to two radiation channels on each side of the slab, while the exterior region of the resonance
band can only couple to a single radiation channel on each side.  

First consider the single-radiation-channel region 
above the dashed purple line in Fig.\ \ref{fig:1}c. Here, there
initially appear to be $4$ unknown parameters in $D = (d_{\textrm{above}},d_{\textrm{below}})^T$, corresponding to two complex outcoupling coefficients to the two radiation channels.
However, as the system is mirror symmetric about $z=0$ and the resonance band's states are even under this symmetry, $D = d_0(1,1)^T$,
with $d_0$ being the remaining complex free parameter.
Moreover, one can show that the constraint represented by Eq.\ (\ref{eq:CDD}) in this region can be written as,
\begin{equation}
  d_0^* (r + t) = - d_0, \label{eq:1rad}
\end{equation}
which amounts to a constraint on the phase of $d_0$, as $|r|^2 + |t|^2 = 1$, where $r$ and $t$ are the direct transmission and
reflection coefficients and $C = (r, t; t, r)$. Thus, in this region there is only a single unknown parameter in $D$, and as there are two degrees of freedom
in the system, $k_x$ and $k_y$, lines of BICs can be found in the resonance band in the single-radiation-channel region along portions
of the edge of the Brillouin zone, as well as near portions of the $\Gamma$--$M$ line, as shown in Fig.\ \ref{fig:1}c.

This same analysis can also be performed in the two-radiation-channel region of the resonance band, closer to the center of the Brillouin zone.
In this region, one finds that $D$ has two unknown parameters, and thus it is possible to find isolated accidental BICs in this region as
there are two degrees of freedom, similar to accidental BICs found in previous works of photonic crystal slabs embedded in homogeneous media \cite{hsu_observation_2013,zhen_topological_2014}. 
For the system shown in Fig.\ \ref{fig:1}a, one finds an accidental BIC near $k_x = 0$, $k_y = 0.2\pi/a$, and a symmetry protected BIC at $\Gamma$
where both radiative channels are even under $C_2$ but the resonance band is odd, as marked in Fig.\ \ref{fig:1}c.

\begin{figure}[t!]
    \centering
    \includegraphics[width=0.45\textwidth]{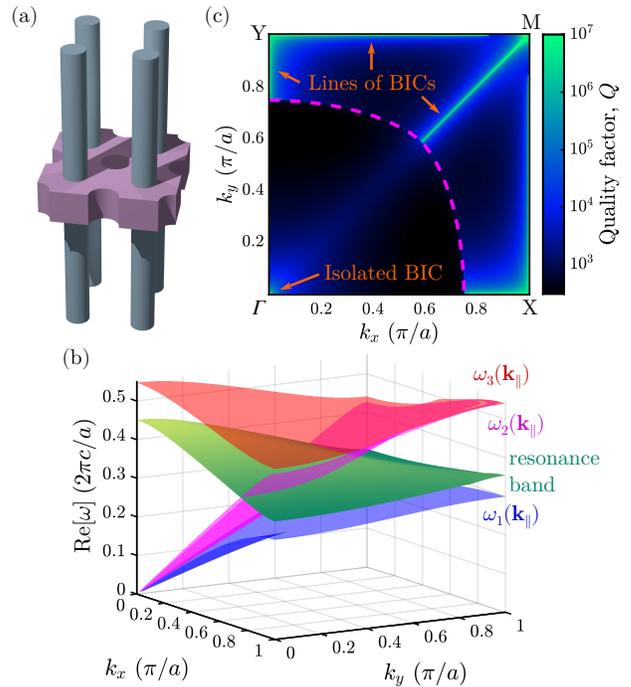}
    \caption{
        (a) Schematic of a photonic crystal slab embedded in a $C_{4v}$ symmetric photonic crystal environment of infinitely tall dielectric rods. The rods have radius
        $r_{\textrm{PhC}} = 0.18 a$ and dielectric $\varepsilon_{\textrm{PhC}} = 12$, while the slab has height $h_{\textrm{slab}} = 0.5 a$, holes with radius $r_{\textrm{slab}} = 0.22a$, and dielectric $\varepsilon_{\textrm{slab}} = 7$,
      where $a$ is the lattice constant of the system. (b) Photonic band structure, with the same conventions as Fig.\ \ref{fig:1}b.
      (c) Quality factor of the photonic crystal slab resonances as a function of $\mathbf{k}_\parallel$. Lines of BICs are seen in the one-radiation-channel region above the dashed purple line, while an isolated BIC is seen at $\Gamma$.
        \label{fig:2}}
\end{figure}

Thus far, we have shown that the environment plays a significant role in determining the distribution of BICs
in a given system. To demonstrate that the environment is an equal partner to the resonant device in determining the presence of BICs,
we increase the symmetry of the photonic crystal environment ($C_{4v}$ as opposed to $C_{2v}$) but preserve the same photonic crystal slab, as shown in Fig.\ \ref{fig:2}a. Thus, although the
spatial profiles of the resonances of the photonic crystal slab remain the same, where they achieve complete destructive interference
in the radiation channels of the new environment has changed, as can be seen by comparing the distribution of BICs found in Fig.\ \ref{fig:2}c,
to the distribution seen in Fig.\ \ref{fig:1}c. In the $C_{4v}$ symmetric environment, the isolated accidental BIC in the two-radiation-channel region
has merged with the isolated BIC at $\Gamma$
and the line of accidental BICs near the $\Gamma$--$M$ line has shifted to lie exactly along $\Gamma$--$M$
and become protected by mirror symmetry about the $x=y$ line.

\begin{figure}[t!]
    \centering
    \includegraphics[width=0.45\textwidth]{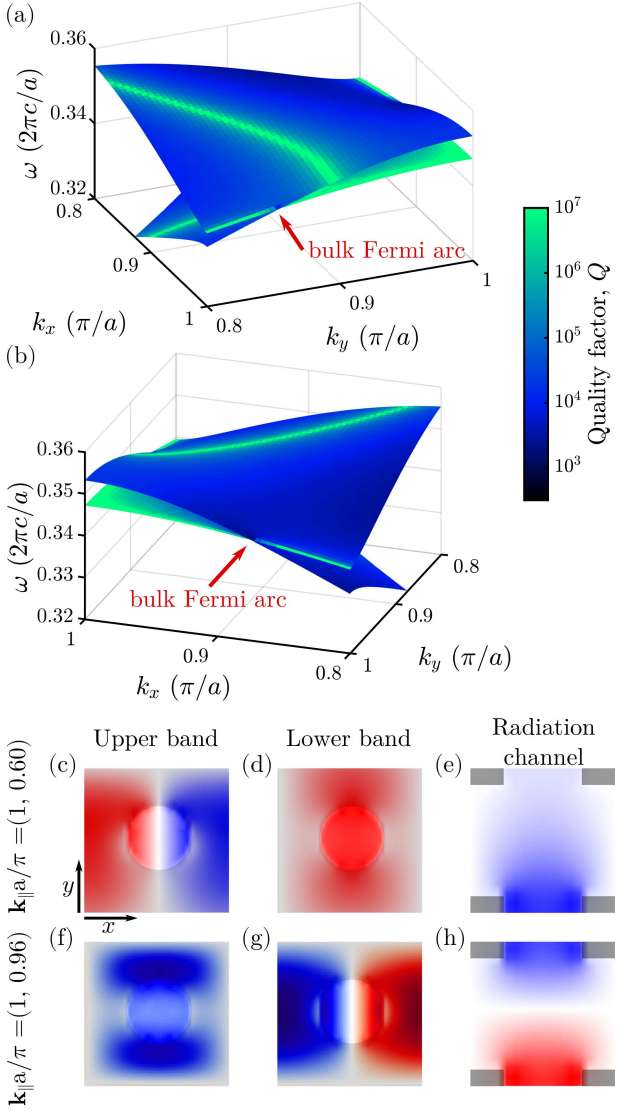}
    \caption{(a)-(b) Frequencies and quality factors near $M$ for two TE-like resonance bands of the photonic crystal slab and $C_{2v}$ symmetric photonic
      crystal environment shown in Fig.\ \ref{fig:1}a. There are two bulk Fermi arcs at $\mathbf{k}_\parallel a/\pi = (1,0.852)$ (marked with a red arrow), and
      $\mathbf{k}_\parallel a/\pi = (0.872,1)$. (c)-(e) Cross-section of the 
      $E_x$ component of the Bloch mode profiles in the $x$-$y$ plane parallel to the photonic crystal slab of the upper resonance band (c), lower resonance band (d), and radiation
      channel (e) at $\mathbf{k}_\parallel a/\pi = (1,0.6)$. Locations of the dielectric features in the photonic crystal slab (c)-(d) or photonic crystal environment (e) are
      denoted with light gray shading. (f)-(h) Similar to (c)-(e), except at $\mathbf{k}_\parallel a/\pi = (1,0.96)$.
      \label{fig:3}}
\end{figure}

There is a curious feature of the lines of BICs along $X$--$M$ and $Y$--$M$ found in both Figs.\ \ref{fig:1}c and \ref{fig:2}c: the lines of
BICs appear to abruptly terminate prior to reaching $M$. Although such a termination is not precluded by the coupled-mode analysis previously discussed,
the line of BICs along $X$--$M$ and $Y$--$M$ are not accidental, but are instead protected by symmetry, as the resonance band is odd about the $x=0$ ($y=0$) plane along this portion of the
$X$--$M$ ($Y$--$M$) high symmetry line, while the radiative channel is even about the same plane, as shown in Figs.\ \ref{fig:3}c and \ref{fig:3}e.
Thus, as the symmetry of the system has not changed at these points along high symmetry lines, it is strange that the modal profile of the resonance band
would suddenly change to allow for the state to couple to the radiative channel.
However, the disappearance of the line of BICs from the resonance band coincides with the location of an intersection with a second
TE-like resonance band of the photonic crystal slab, shown in Fig.\ \ref{fig:3}a and \ref{fig:3}b. Elsewhere in the Brillouin zone these two resonance bands couple and exhibit an avoided crossing,
but along the high symmetry line $X$--$M$ ($Y$--$M$) these two bands have opposite mirror symmetry about the $x=0$ ($y=0$) plane of the system, as
shown in Figs.\ \ref{fig:3}c and \ref{fig:3}d, and thus exhibit a band crossing.

If the coupling to the radiative channels could be ignored such that the system were completely Hermitian, this accidental band crossing would
occur at a Dirac point. Instead, the coupling of the resonance bands to the outgoing radiative channels results in this
system being non-Hermitian. Moreover, 
the two resonance bands in question couple to the single available radiative channel at different rates,
i.e.\ one resonance band possesses a line of symmetry-protected BICs while the other resonance does not. Due to this unequal radiative coupling,
where the hypothetical non-radiating Hermitian system would possess a Dirac point connecting the two bands,
these two resonance bands are instead joined by a bulk Fermi arc in the radiating non-Hermitian system \cite{zhou_observation_2018}.
Bulk Fermi arcs occur generically in non-Hermitian systems and form when a Dirac point is split into two exceptional points \cite{mailybaev_geometric_2005}
connected by a contour where the real part of the frequencies of the two resonance bands are equal, $\re[\omega_+] = \re[\omega_-]$.
When two bands are joined at a bulk Fermi arc, they form two halves of a single Riemann surface.

In the vicinity of the bulk Fermi arc, the effective Hamiltonian for the systems considered here is
\begin{equation}
  \hat{H} = \omega_D - i\gamma + \left(v_{gy} \delta k_y - i\gamma\right) \hat{\sigma}_z + v_{gx} \delta k_x \hat{\sigma}_x, \label{eq:H}
\end{equation}
which results in the spectrum of the resonance bands
\begin{equation}
  \omega_\pm = \omega_D - i\gamma \pm \sqrt{v_{gx}^2 \delta k_x^2 + v_{gy}^2 \delta k_y^2 - \gamma^2 - 2i \gamma v_{gy} \delta k_y}. \label{eq:w}
\end{equation}
Here, $\delta k_x$ and $\delta k_y$ are the wavevector displacements from the underlying Dirac point at $\mathbf{k}_{\parallel,D}$ which
has frequency $\omega_D$, $2\gamma$ is the radiative rate of the resonance band which couples to the single environmental channel,
$v_{gx}$ and $v_{gy}$ are the group velocities describing the dispersion near the Dirac point, and $\hat{\sigma}_{x,z}$ are Pauli matrices.
Equations (\ref{eq:H}) and (\ref{eq:w}) are written for the accidental band crossing along $X$--$M$, but letting $x \leftrightarrow y$
yields the correct set of equations for the accidental band crossing along $Y$--$M$. As can be seen, the spectrum given in Eq.\ (\ref{eq:w})
exhibits a pair of exceptional points at $(\delta k_x,\delta k_y) = (\pm \gamma/v_{gx},0)$, where $\omega_+ = \omega_-$, and which are connected by a
bulk Fermi arc along the contour $\delta k_x < |\gamma/v_{gx}|$ and $\delta k_y = 0$.

With the knowledge that the two resonance bands are connected at a bulk Fermi arc, one can now understand the apparent abrupt termination
of the lines of BICs in Figs.\ \ref{fig:1}c and \ref{fig:2}c.
Along the $X$--$M$ high symmetry lines where $\delta k_x = 0$, one resonance band
remains a BIC with $\im[\omega] = 0$, but the two bands switch when traveling through the bulk Fermi arc, as can be seen
in Fig.\ \ref{fig:3}a. This is because the two connected resonance bands form a single Riemann surface, and trajectories in wavevector space which
pass through the middle of the bulk Fermi arc travel from the upper to lower band, or vice versa.
We can confirm that near $M$ the symmetry of the upper and lower bands along the $X$--$M$ line has switched, such that the lower band exhibits the line of symmetry-protected BICs
while the upper band does not, by 
viewing the modal profiles of the resonances on both sides of the bulk Fermi arc.
As is shown in Figs.\ \ref{fig:3}c-h, for $\delta k_y < 0$, the odd $x$ symmetry mode is found on the upper resonance band,
but for $\delta k_y > 0$ this mode is found on the lower resonance band.
Thus, the symmetry protected
line of BICs does exist along the entire high symmetry line, but passes from the upper resonance band to the lower resonance band through a bulk Fermi arc.

In conclusion, we have demonstrated that the environment within which a device resides significantly constrains
the nature of BICs in different parts of the device's resonance spectrum.
In particular, we have shown that photonic crystal
slabs embedded in a 3D photonic crystal environment can not only exhibit BICs, but can achieve lines of BICs where the environment
only possesses a single radiation channel. This capacity to realize such lines of BICs in photonic structures without birefringent
materials may have applications in on-chip communication and beam steering of photonic crystal surface emitting lasers.
Additionally, we have observed that such simple systems can exhibit highly non-trivial physics,
such as a symmetry protected line of BICs switching from one band of resonances to another by traveling through a bulk Fermi arc.
More broadly though, the ability to engineer the environment rather than the device to realize BICs in a system opens up a broad range
of new experimental possibilities. First, given the advent of advanced 3D-printing techniques such as two-photon
polymerization technology \cite{deubel_direct_2004}, we expect that structures such as the one described here can be straightforwardly fabricated.
However, there are many photonic systems, such as planar grain boundaries in self-assembled structures \cite{blanco_large-scale_2000}, where
controlling the specifics of the embedded device may be very difficult, but engineering the environment is trivial, that
may yield an entirely different route to photonic BICs than has been previously studied.

\begin{acknowledgments}
  The authors acknowledge support from the National Science Foundation under grant numbers
  ECCS-1509546 and DMS-1620422 as well as the Charles E. Kaufman foundation under grant number KA2017-91788.
\end{acknowledgments}


%

\end{document}